\documentclass[prl,reprint,preprintnumbers,nofootinbib,superscriptaddress]{revtex4-1}
\usepackage[titletoc]{appendix}
\usepackage{siunitx}
\usepackage{subfig}
\usepackage{amsmath}
\usepackage{graphicx}

\newcommand{\req}[1]{Eq.\,\eqref{#1}}

\begin{document}

\title{Jet Observable for Photons from High-Intensity Laser-Plasma Interactions}
\author{Scott V. Luedtke}
\affiliation{University of Texas, Austin, Texas 78712, USA}
\affiliation{Los Alamos National Laboratory, Los Alamos, New Mexico, 87545, USA}
\author{Lance A. Labun}
\author{Ou Z. Labun}
\affiliation{University of Texas, Austin, Texas 78712, USA}
\author{Karl-Ulrich Bamberg}
\author{Hartmut Ruhl}
\affiliation{Arnold Sommerfeld Center for Theoretical Physics, Ludwig-Maximilians Universit\"{a}t, Theresienstrasse 37, 80333 M\"{u}nchen, Germany}
\author{Bj\"{o}rn Manuel Hegelich}
\affiliation{University of Texas, Austin, Texas 78712, USA}
\affiliation{Center for Relativistic Laser Science, Institute for Basic Science, Gwangju 61005, South Korea}
\affiliation{Department of Physics and Photon Science, Gwangju Institute of Science and Technology, Gwangju 61005, South Korea}

\date{August 17, 2018}

\begin{abstract}
The goals of discovering quantum radiation dynamics in high-intensity laser-plasma interactions and engineering new laser-driven high-energy particle sources both require accurate and robust predictions.  Experiments rely on particle-in-cell simulations to predict and interpret outcomes, but unknowns in modeling the interaction limit the simulations to qualitative predictions, too uncertain to test the quantum theory.  To establish a basis for quantitative prediction, we introduce a `jet' observable that parameterizes the emitted photon distribution and quantifies a highly directional flux of high-energy photon emission.  Jets are identified by the observable under a variety of physical conditions and shown to be most prominent when the laser pulse forms a wavelength-scale channel through the target.  The highest energy photons are generally emitted in the direction of the jet.  The observable is compatible with characteristics of photon emission from quantum theory. This work offers quantitative guidance for the design of experiments and detectors, offering a foundation to use photon emission to interpret dynamics during high-intensity laser-plasma experiments and validate quantum radiation theory in strong fields.
\end{abstract}

\maketitle

Recent progress in ultra-intense laser physics offers experimental access to a new physics regime, where optical electromagnetic fields are strong enough that particles are accelerated to highly relativistic energies over micron distances and radiate both quantized photons and electron-positron pairs.  
High acceleration, quantum radiation dynamics and particle creation in strong classical fields arise in many areas, e.g., pulsars/magnetars \cite{Harding:2006qn}, black holes \cite{Brout:1995rd}, and the early universe \cite{Parker:1968mv,Anderson:2013ila}, but only laser-matter experiments allow their direct observation in the laboratory and thus validation or falsification of theory \cite{hegelich2017finding}. Ultra-relativistic laser-matter facilities also hold promise for new advanced and compact particle and photon sources suited for applications including radiation therapy, nuclear imaging \cite{jaroszynski2006radiation}, and studies of astrophysics \cite{remington1999modeling}.  Toward both discovery and applications, predictions in some specific configurations suggest that the interaction of a high-intensity ($I\geq 10^{21}\mathrm{W/cm}^2$) laser pulse with an optically-thick plasma produces a high flux of high-energy photons ($E_\gamma\geq 1$ MeV), and engineering the target plasma may control collimation and direction of the photons \cite{ridgers2012,ji2014energy,ji2014near,nerush2014gamma,zhu2015enhanced,stark2016enhanced}. 

However, a long-standing problem has been that predictions are qualitative rather than quantitative.  Although made by extending proven theory of high-energy particle interactions, quantum electrodynamics (QED), into the new regime of strong classical fields, current predictions do not suffice for rigorous discovery, as standard in particle physics.  The challenges of predicting signals of strong-field QED (sfQED) are that relativistic laser-plasma experiments (i) are characterized by both single- and collective-particle motion leading to nonlinear plasma response; and (ii) are difficult to directly observe, requiring femtosecond time resolution and sub-micron spatial resolution in a dense plasma that is opaque to standard probes.  The nonlinear laser-plasma dynamics are investigated by numerically solving the classical, coupled electromagnetic field-plasma equations using kinetic/particle-in-cell methods.  The simulated equations of motion incorporate approximations of the sfQED probabilities for nonlinear Compton scattering off the laser field and Breit-Wheeler pair production \cite{Elkina:2010up,epochQED}.  The difficulty of observing experiments limits the reliability of simulations' predictions, because relevant parameters of the laser-plasma interaction are unmeasured, especially the laser fields on-target and the plasma density profile after ionization and expansion.  Using indirect measurements, such as electron, ion, or photon spectra, requires both quantitatively reliable predictions and systematic comparisons between theory, simulation, and experiment.

To establish quantitative searches for quantum radiation effects in experiment, we introduce a new observable which accounts for the challenges in modeling the laser-plasma interaction.
The observable should isolate high-energy photons, because they are created most copiously at the highest field strength, leave the interaction region immediately, and are not affected by weaker but large spatial extent electromagnetic fields.  It should also show predictable, systematic dependence on input parameters. The simplest, counting photons as a function of angle and energy, is too sensitive to plasma dynamics and shot-to-shot variations, obscuring differences in the laser-plasma dynamics due to systematic changes in experimental parameters, as we will see below.  

We consider the integrated energy flux in high-energy photons in a cone of fixed angular radius $\theta$ and originating at the focal point of laser.  Maximizing this energy over the axis of the cone $\hat\tau$ yields the unique `jet energy'
\begin{align}\label{EJ_theory}
\begin{split}
E_J(\theta) =\underset{\hat\tau}{\rm max}  \int_{E_{cut}}^{\infty}\!\!dE_\gamma\int_0^{\cos(\theta)}\!\! d(\hat k_\gamma\cdot\hat\tau) \oint d\phi \\ \times \int \!\frac{d^3p}{(2\pi)^3}\, f_e(\vec p)\frac{dN_\gamma}{dE_\gamma d\Omega}\vec k_\gamma\cdot \hat\tau,
\end{split}
\end{align}
in which $\vec k_\gamma$ is the photon 3-momentum ($\hat k_\gamma$ the corresponding unit vector), $f_e(\vec p)$ the electron distribution function, and $\Omega$ is the solid angle. The differential emission probability $dN_\gamma/dE_\gamma d\Omega$ can be calculated model-independently and is nonperturbative both because it includes sfQED and because the plasma dynamics modify the classical field during the interaction.  By including only photons above a minimum $E_{cut}\gtrsim 1$ MeV, we separate the large background of low-energy radiation and account for the realistic dynamic range of photon detectors.  Furthermore, considering integrated flux, the detector need not resolve individual photons, which may number in the millions or billions per sterradian presenting a technical challenge--a pile-up problem orders of magnitude worse than the LHC.

To use the jet energy in simulation or experiment, we replace the continuous distributions integrated over in \req{EJ_theory} with sums over particles,
\begin{align}\label{jet_energy}
E_J(\theta)=\underset{\hat\tau}{\mathrm{max}}\sum_{\substack{\hat k_\gamma\cdot \hat\tau>\cos(\theta)\\|\vec k_\gamma|>E_{cut}}}\vec k_\gamma\cdot \hat\tau.
\end{align}
By summing over many (quasi-)particles, computing $E_J$ in a simulation yields a more precise expectation value than simply presenting the distribution of photons in number or energy.   The jet energy is less sensitive to errors introduced by the simulations' approximations of the photon number $dN_\gamma$ \cite{Harvey:2014qla}, because (i) the (average) momentum radiated coincides with the electron momentum even while the emission angle of a single photon may not be accurately reproduced, and (ii) momentum flux suffers no infrared or collinear divergence \cite{Peskin:1995ev,Dinu:2012tj}, whereas the photon number does, making it sensitive to perturbative corrections.  This latter feature makes the jet energy \req{EJ_theory} more amenable systematic improvement by analytic theory calculation. 

In this letter, we present simulations and their analysis using this observable, showing the jet energy can be used to indirectly identify characteristics of the laser-plasma interaction.  Our simulations match parameters of currently available high-intensity lasers.  We perform 2D simulations in the \textsc{PSC} \cite{psc} with a focused, out-of-plane polarized, hyperbolic secant pulse \cite{McDonald:1998pu} with a wavelength $\lambda_l=$\SI{1.058}{\micro\meter}, FWHM pulse duration in intensity of {150} fs, peak intensity of $3.02 \times 10^{22}$ W/cm$^2$ and beam waist radius of \SI{1.25}{\micro\meter}.  The target is a fully-ionized carbon plasma, \SI{10}{\micro\meter} thick except where a preplasma is modeled.  To obtain physical units in our plots, we assume a thickness in the third dimension of \SI{1}{\micro\meter}.  Photon emission is controlled by a Monte Carlo algorithm implementing the differential probability $dN_\gamma/dE_\gamma d\Omega$ in the locally-constant field approximation \cite{Elkina:2010up}.  The code is collisionless and the number of high-energy photons expected from the Bremsstrahlung processes is orders of magnitude smaller than photons contributing to our sfQED emission processes \cite{Wan:2017jhn}.  To ensure reliable simulation results, we average observables over multiple simulations with the same laser and plasma parameters, changing only the initialization of the random number generator (see supplementary information). 

\begin{figure}
\begin{center}	
	\subfloat{\includegraphics[width=0.5\columnwidth]{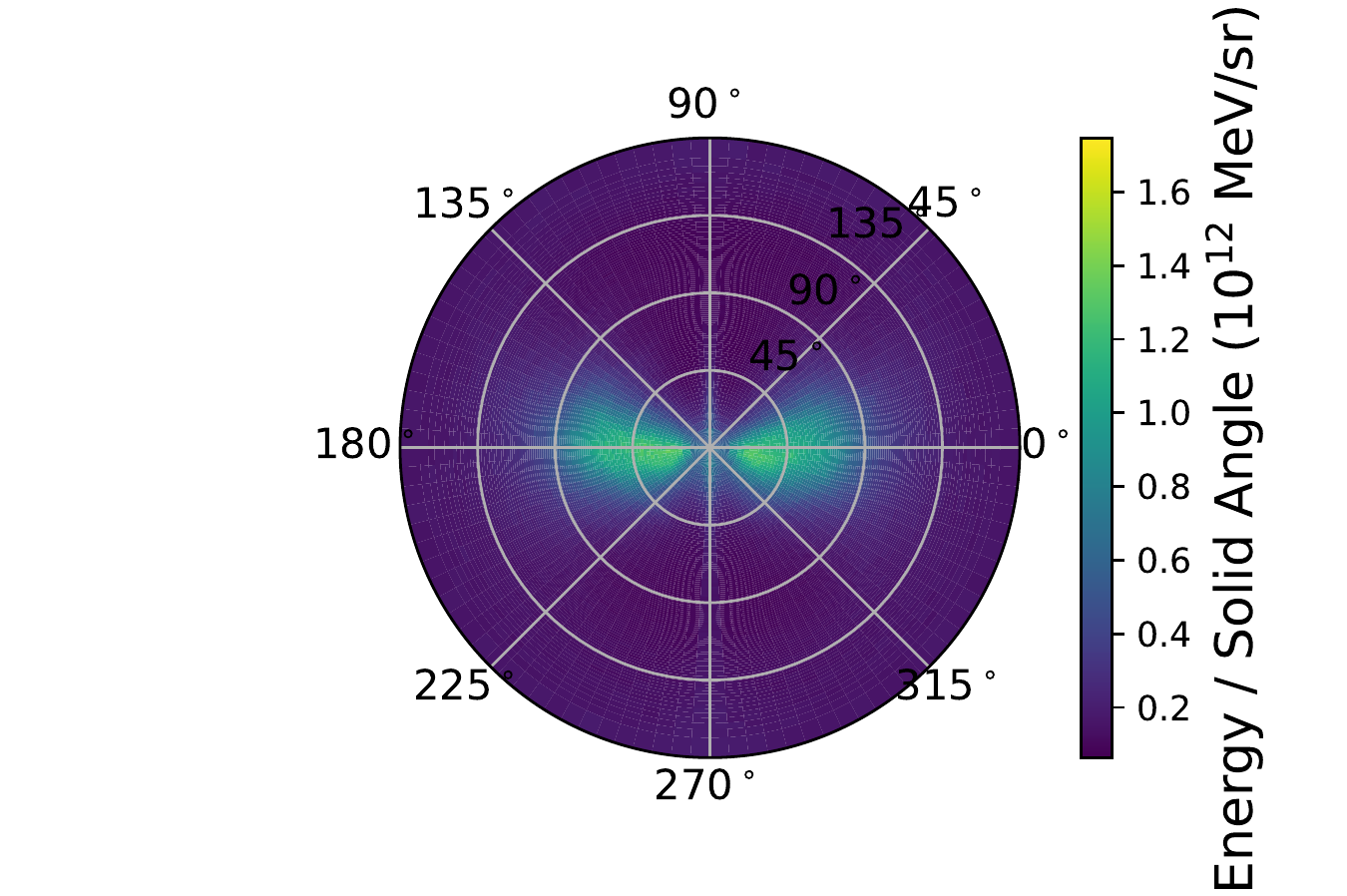}}
	\subfloat{\includegraphics[width=0.5\columnwidth]{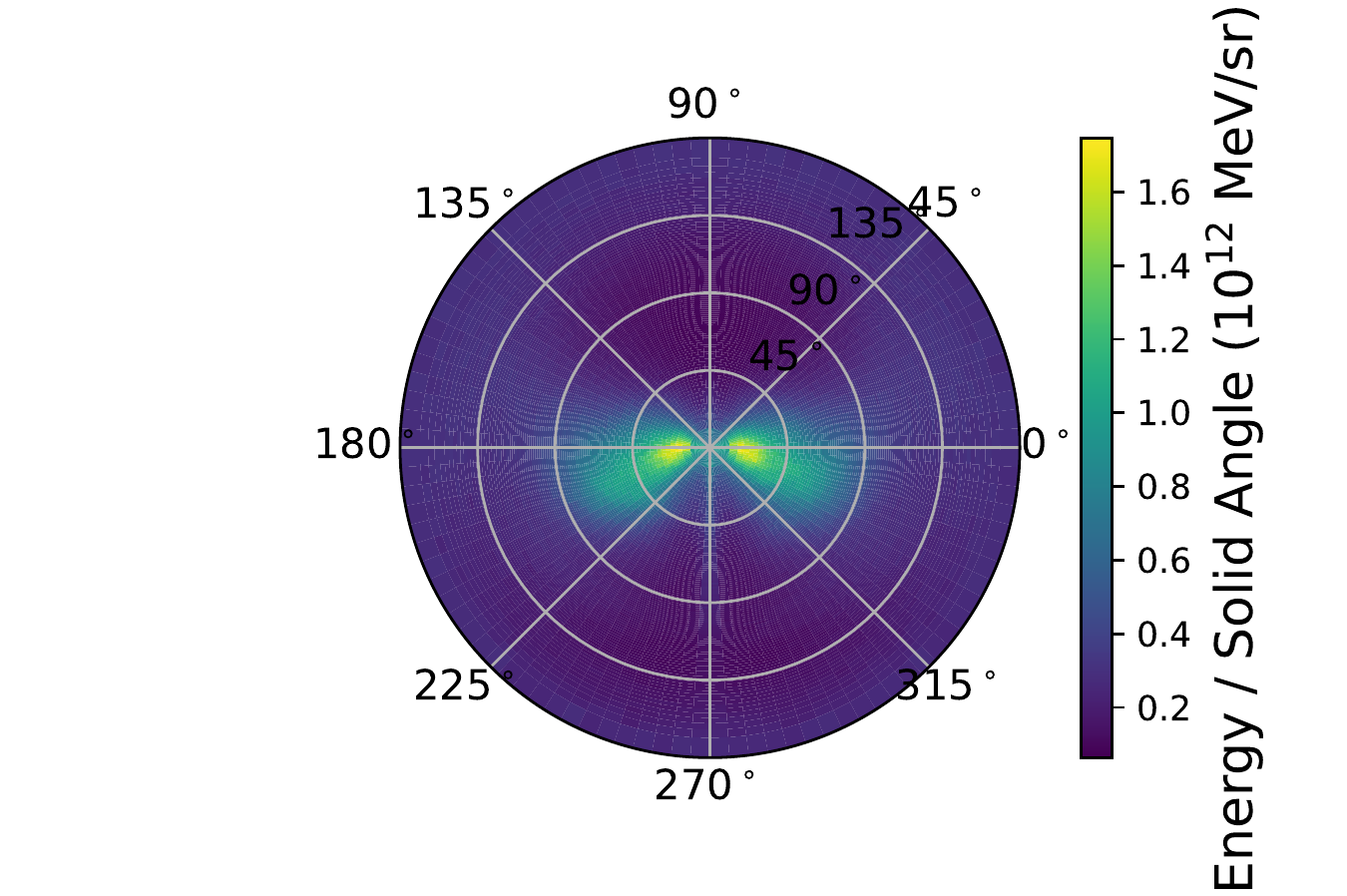}}\\
	\subfloat{\includegraphics[width=0.5\columnwidth]{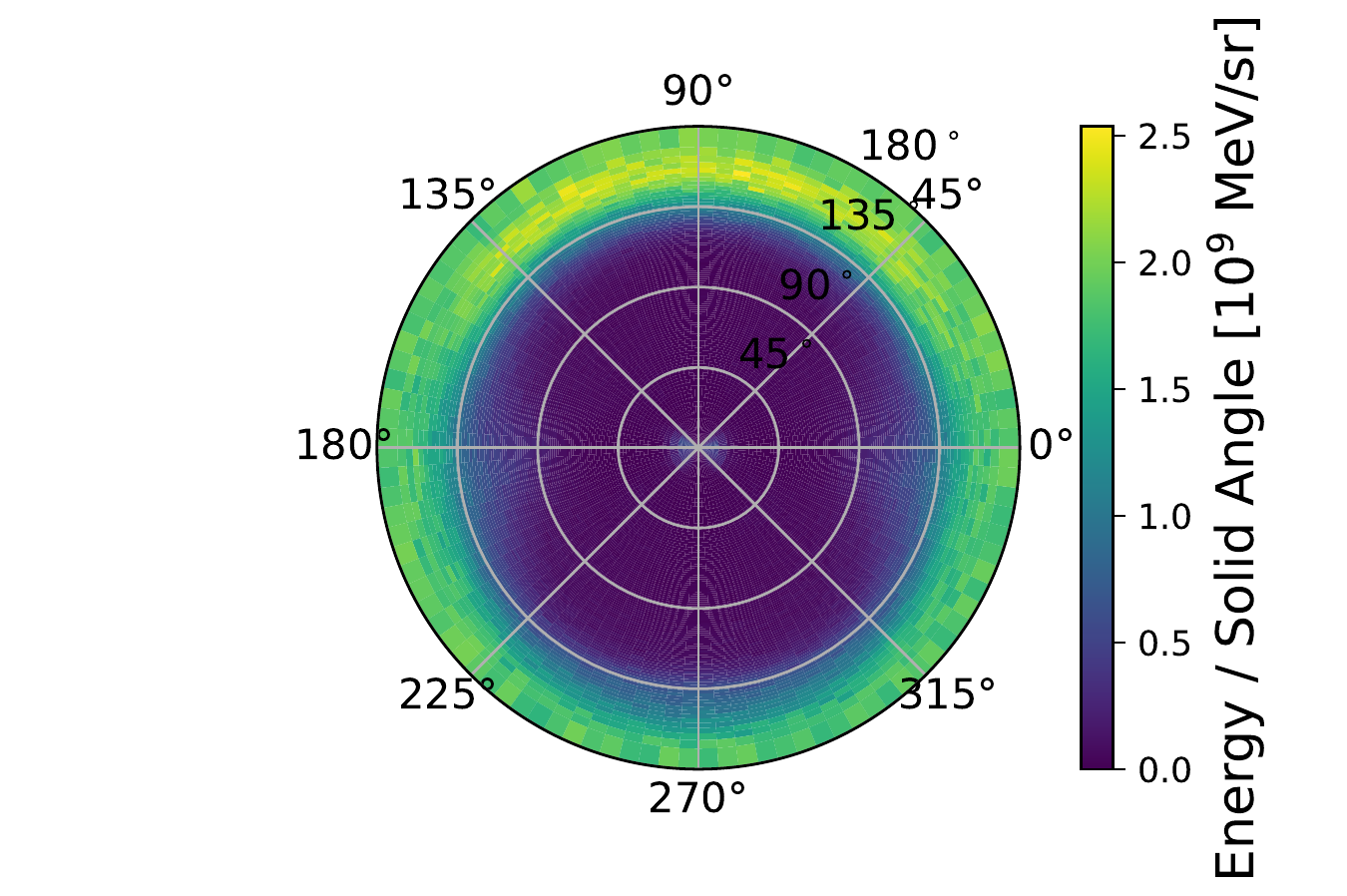}}
	\subfloat{\includegraphics[width=0.5\columnwidth]{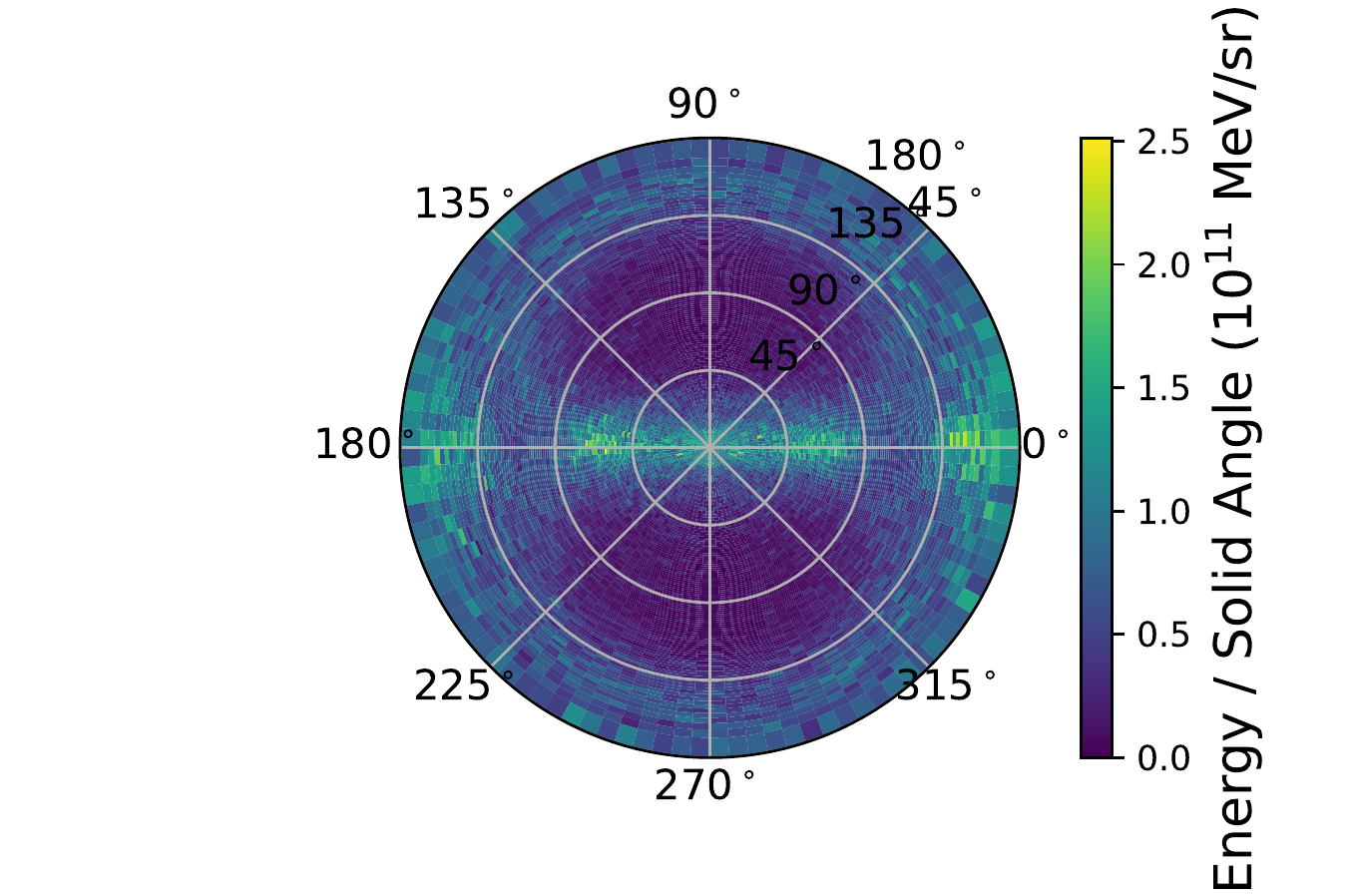}}\\[-6mm]
\end{center}
	\caption[Small multiples]{Angular energy fluxes of high-energy photons ($E_\gamma>1$\,MeV) for simulations of $n_e=60n_{cr}$ slab targets with different random number generator seeds (upper 2 frames), $n_e=3n_{cr}$ (lower left), and $n_e=300n_{cr}$ (lower right).  The laser propagates in the $(0^\circ,0^\circ)$ direction and is polarized along the $\phi=0$ plane. Note the differing color scales for different densities.}%
	\label{fig:cones}
\end{figure}

\begin{figure}
\begin{center}
	\includegraphics[width=0.8\columnwidth]{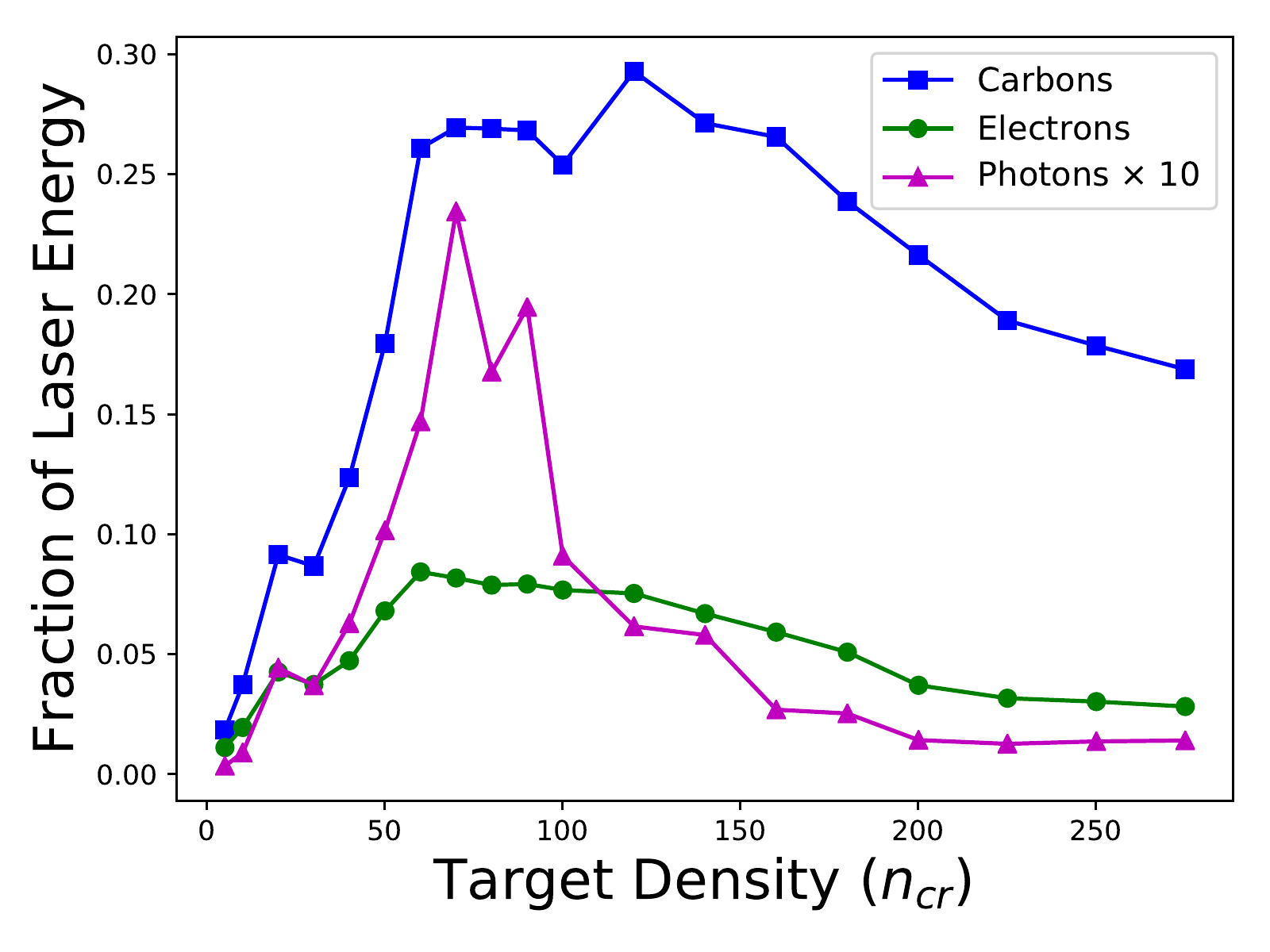}
        \includegraphics[width=0.8\columnwidth]{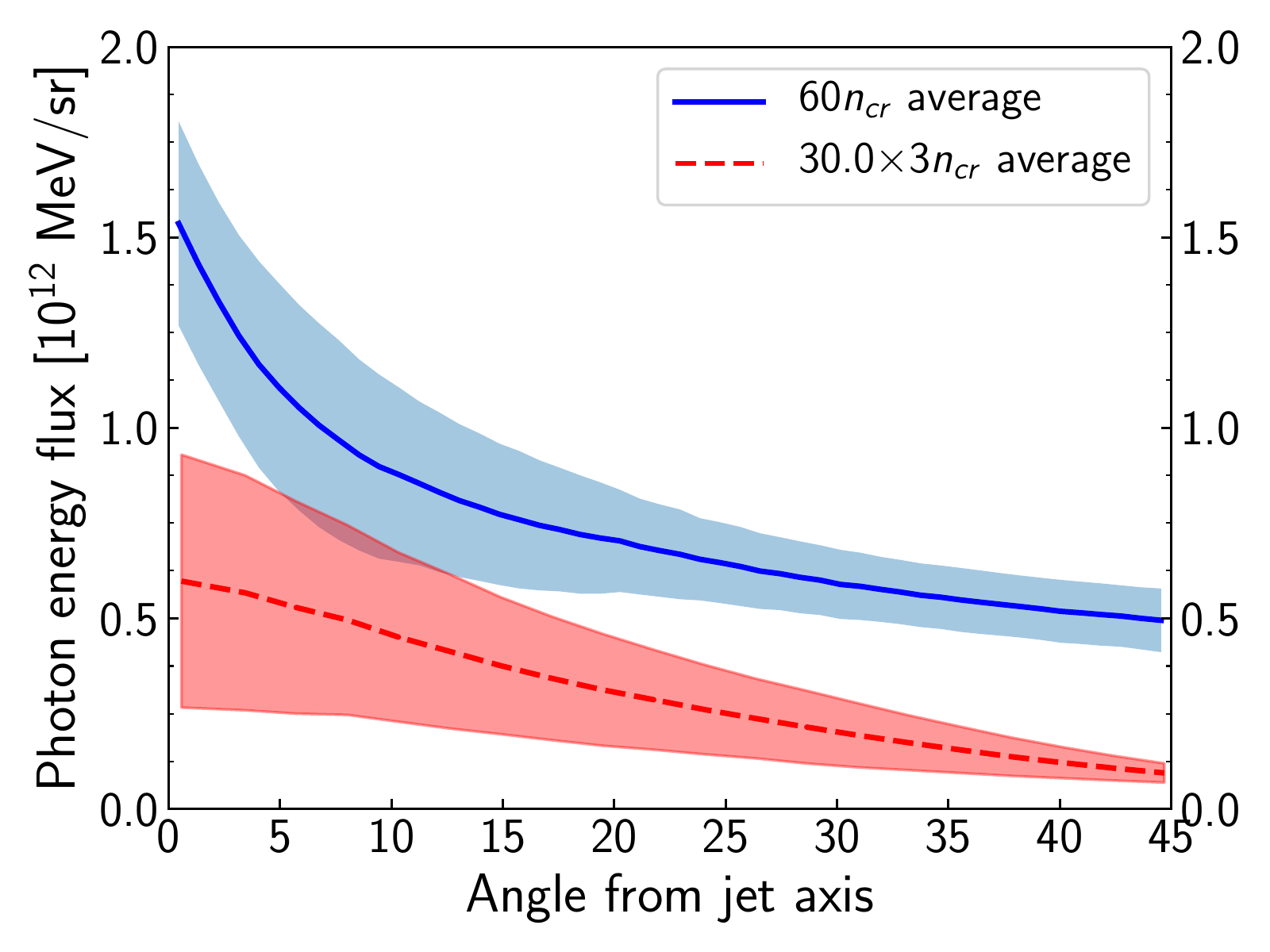}
\end{center}
        \caption[Small multiples]{Top: Fraction of laser energy transferred to each particle species at the end of the simulation as a function of target density.  Bottom: Energy flux in the jet $dE_J/d\Omega$ as a function of angle $\theta$ from the jet axis for $3n_{cr}$ and $60n_{cr}$ slab targets.  The line presents the average and shaded bands the $1\sigma$ variance.  \label{fig:Efrac}}
\end{figure}

The jet energy is especially useful to quantify a characteristic dipolar pattern aligned with polarization of the laser field, seen in examples in Fig.~\ref{fig:cones} (upper panes) and in previous work \cite{ji2014near,nerush2014gamma,stark2016enhanced}. The key feature of the laser-plasma dynamics that enhances total photon emission and the dipolar emission pattern is channeling of the laser pulse in the plasma.  
Photons are collimated into jets by optimizing target density.  Density must be high enough to confine the laser fields $\gg n_{cr}=m_e\omega_l^2/e^2$ ($m_e$ is the electron mass, $\omega_l=2\pi c/\lambda_l$, and $e=|e|$ is the fundamental charge), and low enough to allow the pulse to bore through $<a_0 n_{cr}$ where $(a_0)^2=(I/13.6\mathrm{GW})(\lambda_l/\mathrm{\SI{}{\micro\meter}})^2$ is the (squared) dimensionless intensity parameter \cite{willingale2009characterization}.  Fig.~\ref{fig:Efrac}~(top) shows the fraction of the laser energy transferred to photons is greatest in the range $50n_{cr}$ to~$100n_{cr}$.  

\begin{figure}
    \subfloat{\includegraphics[width=\columnwidth]{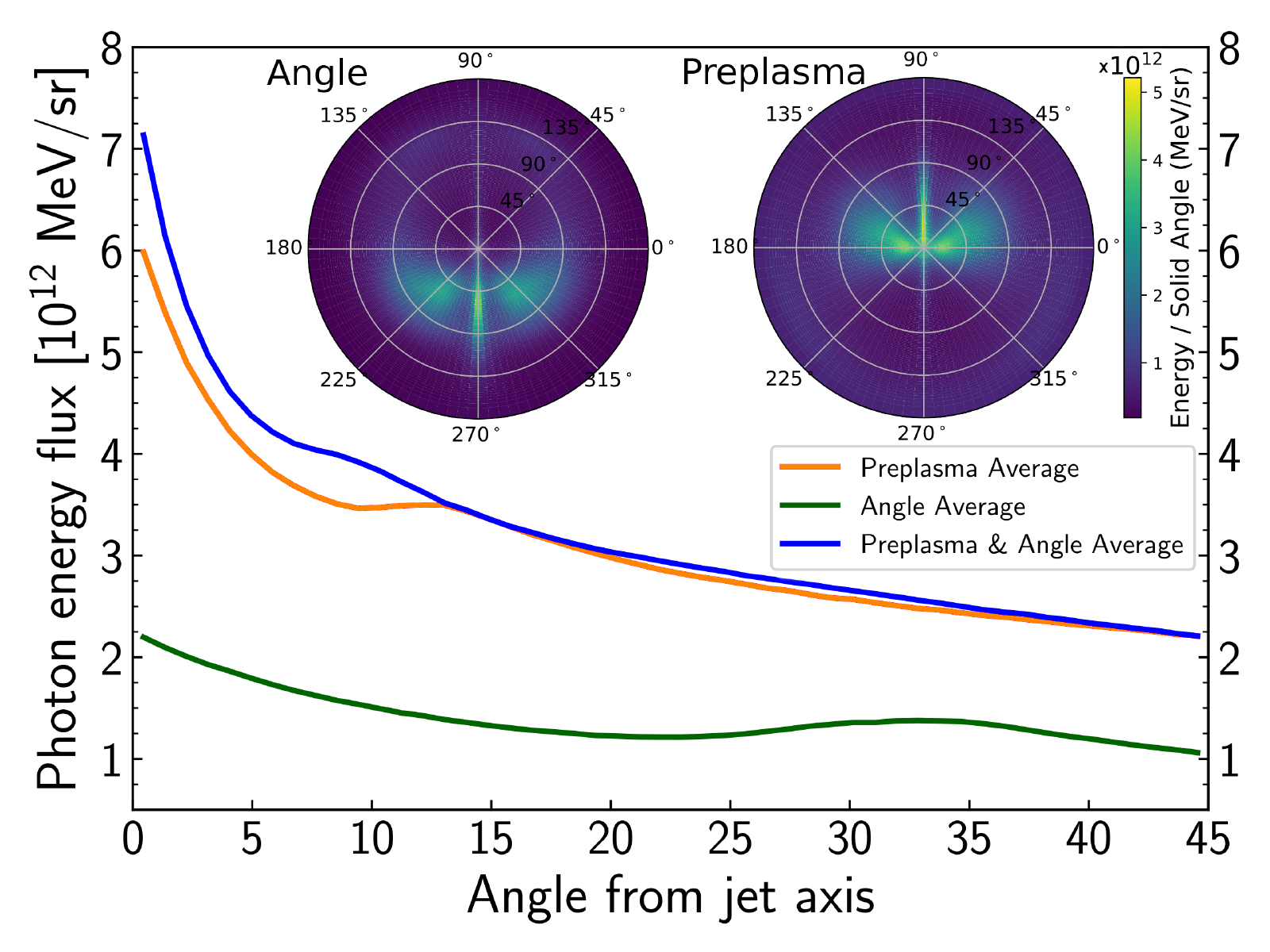}}
	\caption[Small multiples]{Energy flux in the jet $dE_J/d\Omega$ as a function of angle $\theta$ from the jet axis for the simulated experiments with preplasma, $30^\circ$-incident laser, and both.  Lines present averages of two simulations for each.  Insets show example angular fluxes of high-energy photons ($E_\gamma>1$\,MeV) for a laser incident at 30$^{\circ}$ from normal to a $n_e=60n_{cr}$ target (left), and for a normal-incidence laser on a $60n_{cr}$ target with a \SI{5}{\micro\meter}-scale-length preplasma (right).  The bright vertical band is an artifact of the 2D simulation volume and should not be considered physical.  }%
	\label{fig:dEdA}
\end{figure}

Lower density fails to confine the laser to a narrow channel, and the pulse blows through the target transferring little energy to the plasma.  The absolute flux is low and the highest relative flux (Fig.~\ref{fig:cones} lower left) is backward ($\theta\sim 180^{\circ}$), because the emitting electrons are counter-propagating to the laser along the edges of the blown-up channel. 
While the maximization procedure in the jet energy identifies a jet in low-density simulations, it is not statistically significant: the flux on the jet axis differs from other directions by less than 1$\sigma$ (lowest dashed line and band in Fig. \ref{fig:dEdA}).  In contrast, at 60$n_{cr}$, the flux on axis differs by more than $3\sigma$.   Higher density ($\gtrsim 100n_{cr}$) suppresses target deformation and less laser energy is transferred to the plasma, though electrons and ions are heated and accelerated.  Search algorithms fail to identify a unique maximum in $E_J$ to define the jet axis and the flux appears noisy and isotropic (Fig.~\ref{fig:cones} lower right).  Thus the observation of jets (or not) signals a well-formed channel (or not). 

In the optimal density range, the pulse bores through the target in a time of order $L_z/c$, forming a channel of approximately the width of the focal spot $\simeq w_0$.  The channel boundary is charge neutral, displays densities approximately 10 times the initial plasma density and confines the majority of the laser field energy inside.  The laser pulse thus maintains its focus inside the target, and the fields can be enhanced up to four times the input laser intensity $I_0$.  The front of the channel is about half as dense as the walls and is characterized by charge separation. 
Electrons in the front of channel see a sudden increase in field strength as the pulse arrives and begin relativistic oscillatory motion, similar to that of a free electron in a plane wave.  The two-jet pattern arises because the photon emission probability is maximized at the two turning points of the classical trajectory where the electron acceleration is greatest \cite{Ritus:1985}.  Intuitively, this is consistent with the maximum flux of classical radiation.  However, with $a_0\gg 1$, emitted photon energies are up to several hundred times the electron mass, requiring a quantum treatment of the emission process \cite{hegelich2017finding,Labun:2016tbm}. With a large number of electrons in the front of the channel, each with different 3-momentum and emitting at different phases (but most likely near the maximum acceleration), the angular distribution of photons from a single shot is smeared over a finite solid angle compared to the emission predicted from a single electron.

The jet axis $\hat \tau$ correlates to the channel axis, which may deviate from target normal either intentionally or spontaneously. Experimental practice may dictate that the target surface be oriented at an angle to the laser propagation direction in order to suppress retro-reflection, which can damage components up the laser chain.  In this case, the jet axes are offset from target normal and the $\phi=0-\pi$-plane by the laser incidence angle (chosen to be $30^\circ$ for demonstration), seen in Fig.~\ref{fig:dEdA}~(inset left) and Ref.~\cite{ji2014near}.  The on-axis flux increases by roughly the geometric factor $1/\cos(30^{\circ})$. On the other hand, differing jet axes under physically similar initial conditions suggests instabilities in the channeling dynamics.  In Fig.~\ref{fig:cones} (top left), the simulation shows the channel remained near target normal without filamenting, whereas in top right the laser channel deviated below the polarization plane, similar to Figure 1a of \cite{stark2016enhanced}.  In our simulations, the jet axis deviates from the polarization plane by up to \SI{10}{\degree}.  Measuring variance of the jet axis over many shots offers quantitative information about the instabilities.  

In addition to variations in channel direction, the initial plasma density profile may be unknown due to pre-expansion of the plasma, which impacts the laser-plasma dynamics and final particle energies \cite{batani2010prepulses,sentoku2017}.  The energy flux on jet axis increases with either or both angle-incidence and preplasma.  Correspondingly, angled incidence and preplasma transfered about twice the energy to the electrons and carbon ions, and 50\% more to photons than the normal-incidence flat-target simulations.  These outcomes are consistent with the effectively thicker target and so longer channel, as hinted by Ref.~\cite{ridgers2012}.   The shot-to-shot variation in the size and placement of the jets is consistent with variation in the seven normal-incidence simulations.  Thus primary features of the jet energy survive both controlled and uncontrolled variations in conditions, meaning it aids interpretation of experimental results even when simulations do not reproduce precisely the conditions of the experiment.

\begin{figure}
	\centering
	\subfloat{\includegraphics[width=\columnwidth]{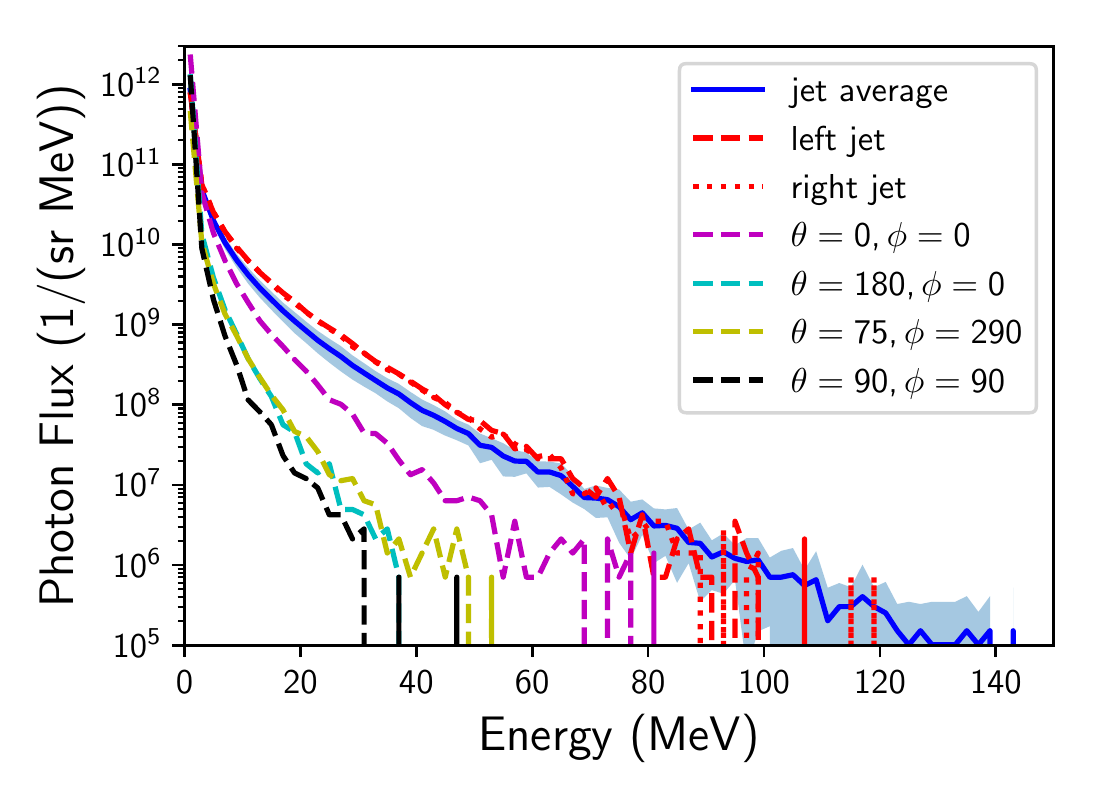}}	
	\caption[Small multiples]{Photon spectra from $60n_{cr}$ target, comparing the jet direction to other directions.  Solid blue line and band indicate average and 1$\sigma$ variation in the jet direction, whereas dashed lines are representative spectra from a single simulation.}%
	\label{fig:spectrum}
\end{figure}

Finally, the distribution of photon energies in the jet carries information about the typical field strengths at emission: the probability to emit at energy $E_\gamma\gg\hbar\omega_l$ is controlled by a power law up to a cutoff frequency \cite{Ride:1995zz,Seipt:2011dx}.  Fig.~\ref{fig:spectrum} shows the photon spectrum $dN_{\gamma}/dE_{\gamma}$ in the jet as compared to other directions for the $60n_{cr}$ target.%  We define the size of the jet shot-to-shot by the angle $\theta$ of steepest descent in energy flux vs. radius.  
The spectra show that both photon number and energy are higher in the jet than in other directions, signaling the highest proper field strengths were seen by electrons emitting into the jet.

These few applications of the jet energy are not exhaustive.  The jet can and should be combined with previous proposed signals \cite{willingale2009characterization} to cross-check and reinforce interpretation of the data in terms of channel-forming.  Though the total energy in photons is only about 2\% of the laser energy, this emission reduces the number and energy of high-energy electrons by an order of magnitude compared to unphysical simulations with sfQED emission disabled.  We expect photon emission to affect particle source applications as soon as detectable numbers of photons are present, and not just when a large fraction of the laser energy is converted to photons \cite{ji2014near}.  To measure radiation losses by the plasma, we should correlate photon observables such as the jet energy with electron observables.  The jet energy also lends itself to defining a quality measure of the photon ``beam'' for applications.

Our results contain lessons for both simulation and experiment.  The jet definition given in \req{jet_energy} is a search algorithm that guides experimental setup and analysis to regions where quantum radiation effects are most apparent. Simulations under idealized conditions better help interpret experimental outcomes when they offer observables that account for and/or smooth over shot-to-shot variations in both the simulation and experiment.  Establishing correlations between measurable features in these observables and the underlying laser-plasma dynamics, such as the examples above for the jet direction and jet shape, can help diagnose experiments.  Experiments should consider new types of detector, particularly calorimeters, that can make useful and reliable measurements in the context of very large particle fluxes.  As our jet-energy example shows, allowing for shot-to-shot fluctuations and model uncertainties in defining the quantities to measure, we can build more systematic and quantitative comparisons of simulations and experiments, leading to improved predictive power and control of outcomes.  

\emph{Acknowledgments.}
Work performed under the auspices of the University of Texas at Austin, the U.S. DOE by Los Alamos National Security, LLC, and Los Alamos National Laboratory . This work was supported by the Air Force Office of Scientific Research (FA9550-14-1-0045) and the LANL ASC and Experimental Sciences programs. High performance computing resources were provided by the Texas Advanced Computing Center. This work used the Extreme Science and Engineering Discovery Environment (XSEDE), which is supported by National Science Foundation grant number ACI-1548562.  We would like to thank B. Albright, D. Stark, and L. Yin for discussions.  O.Z.L. thanks Lynn Ann Labun for her continuous and generous support.

\appendix
\section{Appendix A: Simulation Details}

We run 2-dimensional PIC simulations in the \textsc{PSC} \cite{psc}.  The laser pulse is the same in all simulations: a focused, polarized out of the simulation plane \cite{stark2017effects} pulse with a wavelength of \SI{1.058}{\micro\meter}, 150 J total energy, FWHM pulse duration in intensity of \SI{150}{\femto\second}, peak intensity of $3.02 \times 10^{22}$ W/cm$^2$, beam waist of \SI{1.25}{\micro\meter}, and hyperbolic secant temporal profile.  The total simulation time is \SI{150}{\pico\second}.  The pulse is initialized with the peak 533~fs from the front face of the target, far enough to ensure that the laser electric field strength is the same order of magnitude as root-mean-square electric field due to thermal fluctuations in the plasma.  

The plasma is a \SI{10}{\micro\meter}-thick slab of electrons and fully-ionzied carbon (C$^{6+}$).  The preplasma is modeled with a Gaussian envelope with a standard deviation of \SI{5}{\micro\meter}, extending \SI{10}{\micro\meter} in front of the laser focus, such that the total mass of the target is conserved.  In the angle-incidence simulations, the laser wave vector is aligned \SI{30}{\degree} from the target normal.

We check convergence by varying the number of particles per cell, the initial electron temperature and grid cell size.  With the photon-emission model disabled, a range of quasi-particle numbers from as few as 12$e^-$/2C$^{6+}$ per cell to 180$e^-$/60C$^{6+}$ all yield consistent results for plasma dynamics and particle observables.  The results analyzed for jet energy and photon spectra use simulations with 60$e^{-}$/20C$^{6+}$ per cell for higher-density targets ($n_e= 60n_{cr},300n_{cr}$) and both 60$e^{-}$/20 C$^{6+}$ and $35e^{-}$/10 C$^{6+}$ per cell for the $n_e=3n_{cr}$ target.  The ion temperature is 10 eV and the electron temperature is 10 keV, above which the energy transferred to the electrons increases in variability but changes little in average.  Grid cell sizes \SI{20}{\nano\meter} or smaller yield consistent averages for particle and photon observables.  The presented results use simulations with cell sizes of 20, 10, and \SI{5}{\nano\meter}.

To ensure reliable simulation results, we run several simulations with the same simulation parameters, changing only the initialization of the random number generator.  As the random number generator affects both initial quasi-particle positions and the stochastic emission of photons, observables from these simulation sets can be averaged to yield a better estimate of the observable's expectation value.  Further, the two jets within an experiment are indistinguishable, because the reflection symmetry in the laser electric-field axis is broken only by the carrier-envelope phase, which is not controlled in current experiments.   Correspondingly, the two jets in the same simulation are very similar, never differing by more than a few percent in any metric we used.  Therefore we average the jet energy and spectrum over the two jets within each simulation.  For the same reasons, the standard deviation in $dE_J/d\Omega$ is computed by first averaging over the two jets in a single simulation and then then root-mean-square deviation from the average over the multiple simulations.  The number of simulations contributing to the statistics in the figures in the text is given in Table~\ref{tab:numsims}.

\begin{table}
\begin{tabular}{lc}
\hline \hline Simulation (density/modifications) & Number of simulations \\ \hline
$3n_{cr}$ & 9 \\
$60n_{cr}$ & 7 \\
$60n_{cr}$+preplasma & 2 \\
$60n_{cr}$+($30^\circ$ incidence) & 2 \\
$60n_{cr}$+($30^\circ$ incidence)+preplasma & 2 \\
\hline \hline
\end{tabular}
\caption{Number of physically identical simulations averaged into the jet observable data presented in Figs.~\ref{fig:Efrac}~(bottom), \ref{fig:dEdA}, and \ref{fig:spectrum}.\label{tab:numsims}}
\end{table}

\section{Appendix B: Jet-finding algorithm}

The jet axes were found using SciPy's Nelder-Mead optimization algorithm \cite{nelder1965simplex,wright1996direct} maximizing Eq.~\ref{jet_energy} over the ordered pair (a spherical angle)
composing the jet axis, $\hat \tau$.   Nelder-Mead used many fewer function calls than other methods, converging in a similar number of iterations without computing derivatives.  We choose a jet angular radius of 0.15~radians, which is small enough to exclude other features from the jet and big enough to include enough simulation photons for a smoothly varying jet energy.  We tested various tolerances for the angles at various seed angles and found that below \SI{1}{\degree} tolerance, the standard deviation of resulting angles was stable, with $0.1<\sigma<1$, and with the average angles agreeing within the greater of \SI{1}{\degree} or the tolerance.  Therefore, we used a tolerance of \SI{1}{\degree}, and conclude that our jet axes are accurate to $\pm$~\SI{1}{\degree}.  We choose seed angles for the optimization manually, and checked the result visually.  We occasionally re-chose the seed angles to converge in the correct jet.  For the 300$n_{cr}$ simulations, the data are sparse and noisy, and the algorithm does not converge to a jet axis that passes visual checks except trivially by seeding the algorithm with what looks like a jet axis. (See Fig.~\ref{fig:cones}f.)

\bibliographystyle{apsrev4-1}
\bibliography{Jet_Observable}

\end{document}